\documentclass[draft]{agujournal2019}
\usepackage{url} 
\usepackage{lineno}
\usepackage{amsmath}
\usepackage[capitalise]{cleveref}
\usepackage[normalem]{ulem}
\usepackage[inline]{trackchanges} 
\usepackage{soul}
\soulregister\cite7

\draftfalse

\journalname{JGR: Planets}

\begin{document}

\title{The Length of Martian Crater Rays and Their Relation to Lunar Cold Spots}

\authors{Trevor P. Erwin\affil{1}, Brandon C. Johnson\affil{1,2}, David Minton\affil{1,2}, Alexandria V. Johnson\affil{1}}

\affiliation{1}{Department of Earth, Atmospheric, and Planetary Sciences, Purdue University, West Lafayette, IN 47907}
\affiliation{2}{Department of Physics and Astronomy, Purdue University, West Lafayette, IN 47907}

\correspondingauthor{Trevor Erwin}{erwin19@purdue.edu}

\begin{keypoints}
\item Martian crater rays are longer than lunar crater rays by an order of magnitude
\item Lunar cold spots have ray lengths similar to Martian crater rays
\item We propose a relationship between the formation mechanisms of lunar cold spots and Martian crater rays
\end{keypoints}

\begin{abstract}
Impact-generated crater rays are well-documented on the Moon, with most appearing as high-albedo streaks extending radially from a crater’s center. On Mars, however, crater rays are significantly rarer and discernible only through thermal imaging due to their lower thermal inertia compared to surrounding terrain. This study presents the first comparative analysis between the lengths of Martian and lunar crater rays, including lunar albedo rays and cold spots, which are ray-like thermal anomalies associated with many of the youngest lunar craters. Our findings indicate that both Martian crater rays and lunar cold spots extend significantly farther than lunar albedo rays, with lengths an order of magnitude greater for craters of equivalent diameter. Furthermore, we propose a connection between the formation mechanisms of Martian crater rays and lunar cold spots based on their thermal properties. By integrating thermal rays into existing ejecta models, we refine the understanding of crater-ray formation and suggest that Martian crater rays and lunar cold spots may share a similar formation mechanism via secondary cratering processes. Advancing knowledge of these features has implications for impact dynamics and surface evolution across planetary bodies.

\end{abstract}

\section*{Plain Language Summary}
Craters on both the Moon and Mars can form rays, which are streaks extending radially from a crater's center. Lunar rays are much more common, and often have bright albedo features, appearing lighter than surrounding material. Martian rays, however, are much rarer, and are only visible in thermal imaging due to their low thermal inertia compared to background material. This work compares the lengths of Martian and lunar crater rays for the first time, as well as lunar cold spots, which, like Martian crater rays, are often only visible using thermal imaging due to their lower thermal inertia. This work finds that both Martian crater rays and lunar cold spots are longer than albedo lunar rays by an order of magnitude for craters of the same diameter. In addition, we propose that the mechanisms for the formation of Martian crater rays and lunar cold spots may be linked due to their unique thermal properties. Properly understanding the formation of these thermal rays will lead to a better understanding of impact dynamics, and could help to redefine how we measure ejecta patterns on different bodies.
\newpage

\section{Introduction}
Crater rays are a common feature of rocky bodies in our solar system. These rays are streaks of material with bright albedo features that extend radially from fresh craters \cite{Melosh1989}. Their lengths vary, but longer ones, such as those surrounding Tycho crater on the Moon, span nearly an entire hemisphere \cite{Baldwin1963}. Crater rays are observed on the Moon \cite{Baldwin1963}, Mercury \cite{Neish2013}, Mars \cite{McEwen2005}, and the moons of Jupiter \cite{Schenk1991}. Before their discovery on Mars by \citeA{McEwen2005}, it was widely believed that atmospheric processes would prevent crater ray formation \cite{Melosh1984}. However, more than 100 rayed craters have since been identified on the Martian surface \cite{Harris_2023}, though unlike their lunar counterparts, Martian crater rays are primarily visible in thermal data \cite{McEwen2005}. 

Lunar rays have been classified into two categories; immature rays and mature rays \cite{Hawke2004}. Immature rays form immediately upon impact, consisting of freshly excavated high-albedo material that creates bright streaks across the lunar surface. Over time, space weathering darkens these rays, transitioning them into mature rays. Mature rays are often darker, but some can sill appear bright due to compositional differences, where the ejected material has a higher intrinsic albedo than the surrounding terrain. These mature rays will eventually darken to the point where they are no longer distinguishable from the background regolith. Viewing conditions also affect ray visibility, with low-incidence angles enhancing contrast on both the Moon and Mercury \cite{Neish2013}. 

\begin{figure}
    \noindent\includegraphics[width=\textwidth]{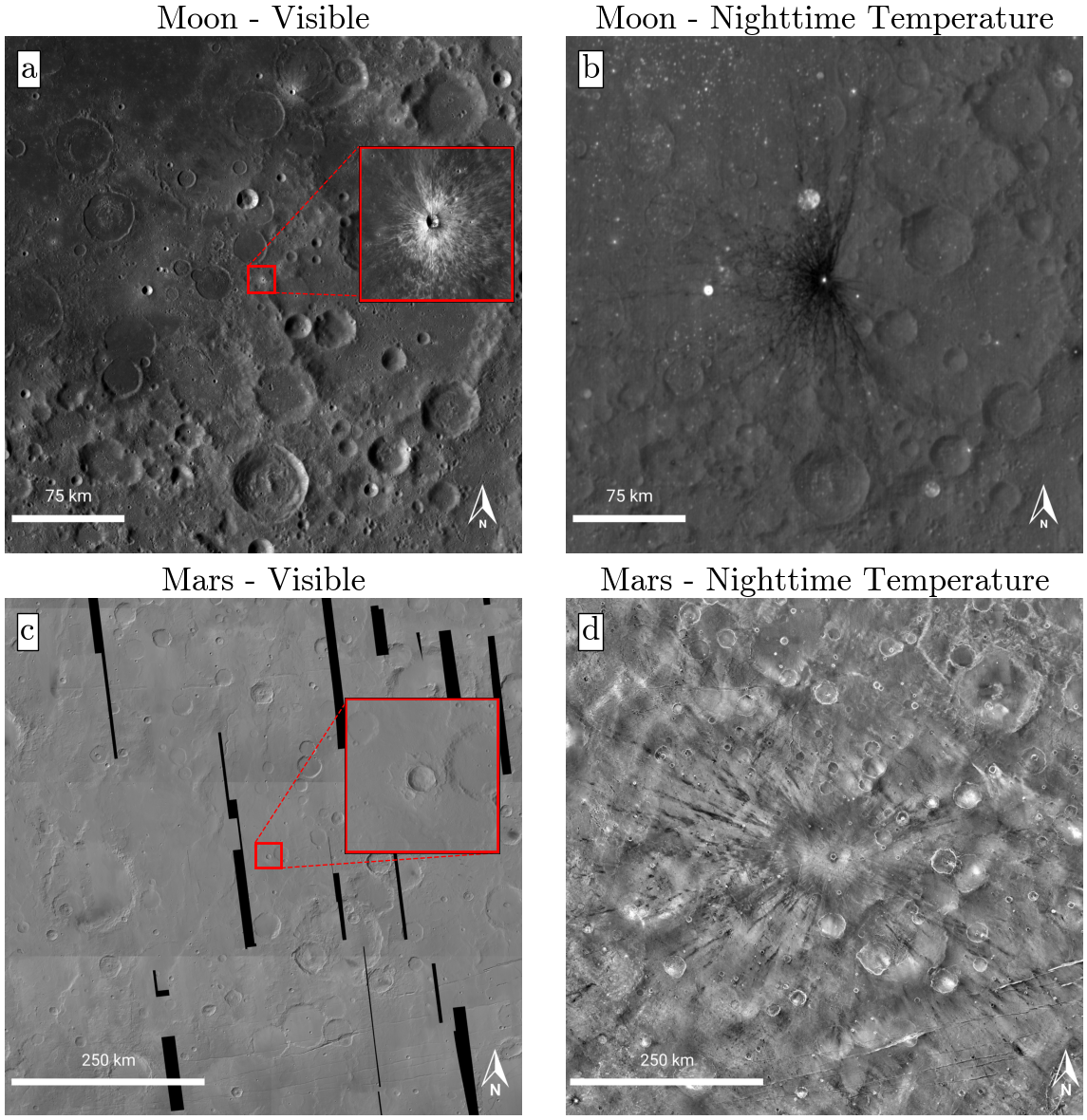}
    \caption{(a) Bandfield crater, a 0.9 km diameter lunar crater located at 5.39°S, 90.76°E. The crater itself is enlarged to show the lengths of the visible crater rays. (b) Nighttime rock-free regolith temperature deviations of the Bandfield crater using LROC WAC global mosaic for shading showing a visible cold spot. Darker portions of the image have a lower surface temperature. (c) Gratteri crater, a ~6.9 km diameter Martian crater located at 17.7°S, 199.9°E. As in \cref{fig:Bandfield_ColdSpot}a, the primary crater is enlarged, but the Martian rays are not visible to the naked eye. (d) Nighttime temperature measurements of Gratteri crater using THEMIS. Darker portions indicate areas with lower temperature reading.   \textit{Figures adapted from \citeA{Williams_2018} and \citeA{Tornabene2006}}.}
    \label{fig:Bandfield_ColdSpot}
\end{figure}

Martian crater rays exhibit notable differences from their lunar counterparts. Mars hosts significantly fewer rayed craters \mbox{\cite{Tornabene2006}}, and its rays are detectable only in thermal infrared wavelengths due to their intrinsically low thermal inertia (see \mbox{\cref{fig:Bandfield_ColdSpot}}d). Unlike the Moon, where impact gardening dominates surface modification, Martian rays are more rapidly eroded by aeolian processes, resulting in a population of comparatively younger rayed craters \mbox{\cite{Tornabene2006}}. Because these rays manifest solely as streaks of low thermal inertia, they are difficult to distinguish in terrains that already possess low thermal inertia. Consequently, Martian rays are primarily observed within young volcanic plains, where the medium to high background thermal inertia enhances their detectability \mbox{\cite{Tornabene2006,Preblich2007, Mellon_2000}}. Despite these differences, both lunar and Martian rays consist of chains and clusters of secondary craters with overlapping ejecta \mbox{\cite{Tornabene2006}}.

Thermal rays, however, are not unique to Mars. Thermal mapping of the lunar surface using the Lunar Reconnaissance Orbiter (LRO) has revealed similar features surrounding lunar craters known as ``cold spots". Analysis of LRO Diviner Lunar Radiometer Experiment data by \citeA{Bandfield_2014} identified these regions as having distinct nighttime temperature drops of up to 10 K. Cold spots exhibit rayed profiles that mimic crater rays, and have lower thermal inertia than their surroundings \cite{Hayne_2017}. \cref{fig:Bandfield_ColdSpot}b, adapted from \citeA{Bandfield_2014}, demonstrates the radial symmetry of these cold spots. The formation mechanism of cold spots remains uncertain, though several theories exist, including the idea that an impact can transmit a wave through the ground to ``fluff up" the surface material \cite{FRIZZELL_2025}. Recent studies show that all new lunar craters develop cold spots, where their diameter and nighttime temperatures scale directly with impactor size \cite{2021AGU_Powell}. Additionally, cold spots extend far beyond the lengths of lunar rays from the same crater (\citeA{Ghent_2024} and this work).

Rayed craters play a crucial role in remote sensing, helping to establish a chronological framework for planetary geology. While they are not reliable indicators of absolute crater age \cite{Hawke2004}, their presence or absence can be used to determine relative ages of nearby features.  Understanding their formation is essential for assessing how they interact with the surrounding terrain, particularly in identifying correlations between the primary crater radius and ray length. \citeA{Elliott2018} demonstrated that lunar crater ray length is controlled by the depth of excavation by secondary craters relative to the space-weathered upper regolith layer. This study extends their methodology to Martian craters, which share the same secondary cratering formation process \cite{Tornabene2006}. 

In Section 2, we expand on the work done by \citeA{Elliott2018} - applying their model of secondary ray formation to Martian crater rays. Section 3 examines lunar cold spots, and the similarities in formation mechanisms and thermal inertia. The minimum excavation depth needed for ray formation on each body is then discussed in Section 4. Finally, we discuss the similarities between Martian crater rays and lunar cold spots and explore potential connections between their formation mechanisms.

\section{Constraining Martian Rays}

\subsection{Crater Ray Modeling}
\begin{figure}
    \noindent\includegraphics[width=\textwidth]{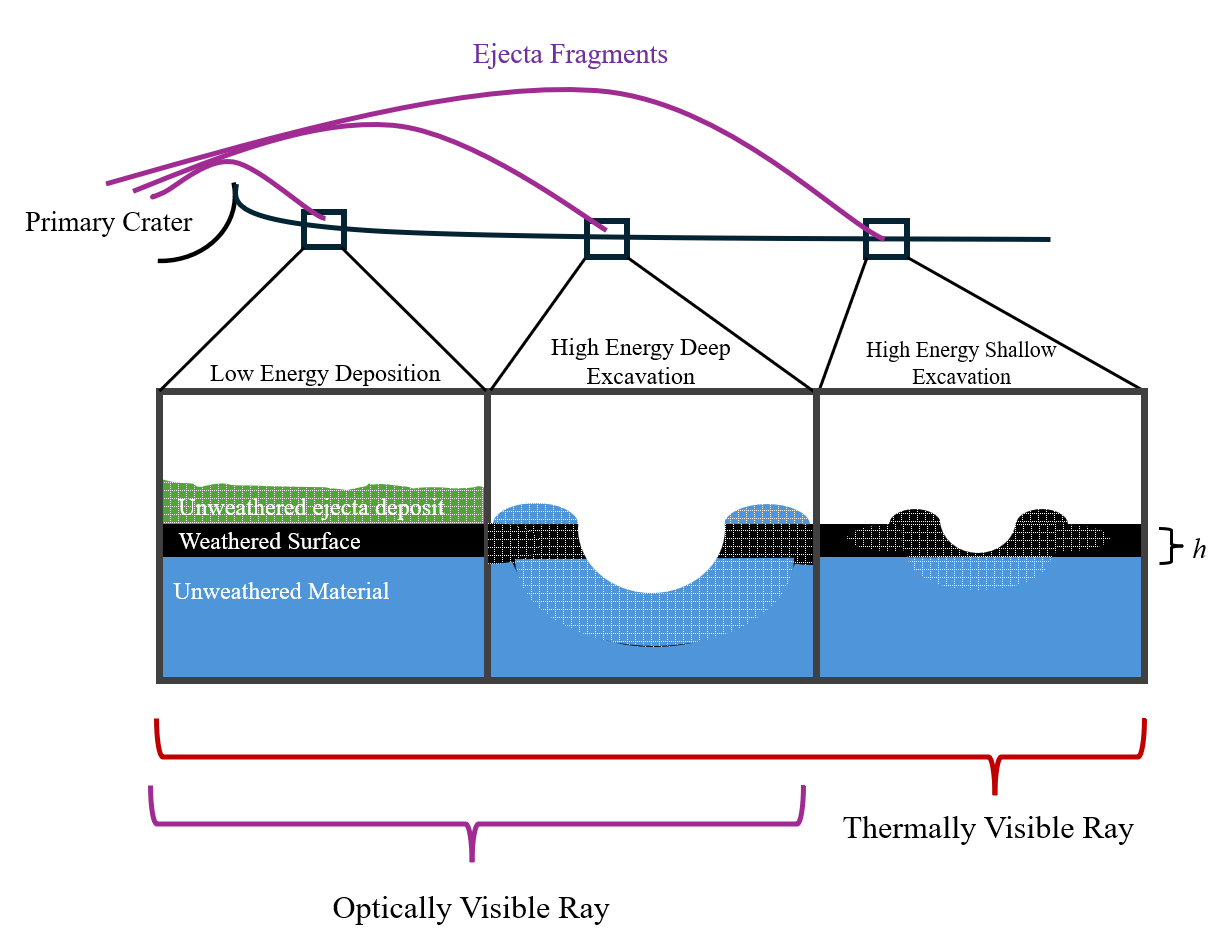}
    \caption{Different crater ejecta regimes proposed by \citeA{Elliott2018}. The first, low energy deposition, happens closest to the crater rim. It forms when the ejected mass has a low enough velocity that it does not mix with surface material, forming a bright layer of ejecta on top of the pre-existing darkened surface. The second, high energy deep excavation, occurs via secondary cratering, where ejected material has enough mass and velocity to excavate a depth of $h$ through the surface to create visible regions of bright unweathered material. The last regime, high energy shallow excavation, is still caused by secondary cratering, but at distances very far from the crater rim where the ejecta does not have enough mass to excavate a depth of $h$, so the ray appears to end. Although work by \citeA{Elliott2018} defined the ray as only being optically visible in the first two regimes, our work indicates that the ray remains thermally visible across all regimes. Likely caused by a change in the porosity of the material (represented by the hatched regions) around the secondary impactor in the latter two regimes. \textit{Figure adapted from \citeA{Elliott2018}.}}
    \label{Ejecta Types Elliott}
\end{figure}

This study adapts the methodology from \citeA{Elliott2018} to compare Martian rays and lunar cold spots. Ray mapping in \citeA{Elliott2018} was conducted using JMARS (Java Mission-planning and Analysis for Remote Sensing), a geospatial software designed for planetary studies \cite{Dickenshield_2015}. Trends observed in the visual data were then explained through secondary crater scaling. \citeA{Elliott2018} proposed that following the initial impact, the ejecta curtain launches large boulders that generate secondary craters, and used boulder density mapping around Censorinus crater from \citeA{Krishna2016} to support this idea.  This framework distinguishes three crater ray regimes; Low Energy Deposition, High Energy Deep Excavation, and High Energy Shallow Excavation, described in further detail in \cref{Ejecta Types Elliott}. According to \citeA{Elliott2018}, rays are only visible in the first two regimes, as the third regime lacks sufficient energy to excavate the ground at depth of $h$ and expose bright material.

\citeA{Baldwin1963} first identified a relationship between primary crater diameter and ray length using photometric data from \citeA{kuiper1960}, suggesting that the ray length was proportional to crater diameter. Work by \citeA{Moore1974} later demonstrated that a power-law was a better fit for Baldwin's data, yielding a refined relationship for lunar craters defined as:
\begin{equation}
    R_r=5.25R_p^{1.25}
    \label{Moore_Equation}
\end{equation}
where $R_r$ is the length of the crater rays and $R_p$ is the radius of the primary crater.

\citeA{Elliott2018} expanded on the work by \citeA{Moore1974} by deriving mathematical relationships for the two constants, referring to them as $R_1$ and $s$ respectively (see \cref{Elliot_equation} below).

The formula derived by \citeA{Elliott2018} is: 
\begin{equation}
    R_r=R_1R_p^s,
    \label{Elliot_equation}
\end{equation}
where 
\begin{align}
    R_1&= \left(\frac{h}{2\alpha}\right)^{\frac{-2-b/(1-\beta)}{a-\beta b/\left(1-\beta\right)}}\left|\frac{\ln{(1-P)}}{-c_2}\right|^{\frac{1}{a-\beta b/\left(1-\beta\right)}},\label{R1 equation}\\
    s&=\frac{a-c}{a-\beta b/(1-\beta)}, \label{s equation}
\end{align}
and $c_2=\frac{c_1}{\left(1+\frac{b}{1-\beta}\right)\left(2+\frac{b}{1-\beta}\right)}$ and $c_1$= $\pi n_0 d_{boulder}^{-b} k^\frac{b}{1-\beta} \left(\frac{\delta}{\rho}\right)^\frac{-b}{3\left(1-\beta\right)} R_{ref}^{-c} |\frac{b}{1-\beta}|$. For the full derivation of these equations, refer to section 4 of \citeA{Elliott2018}. 

The primary use of the fit derived by \citeA{Elliott2018} is to constrain the minimum space-weathered skin depth $h$ needed to form a lunar crater ray (see \cref{Ejecta Types Elliott}), where a majority of the parameters from \cref{R1 equation} and \cref{s equation} can be constrained from observational data (see \cref{Table:Parameters} in the Appendix for a list of these values and their descriptions). In this work, we adopt the same parameter $h$, but generalize its interpretation: rather than tying it specifically to space weathering, we use $h$ to represent the minimum excavation depth required for ray formation across planetary bodies. This distinction is important because, unlike lunar albedo rays, Martian crater rays and cold spots are not governed by space-weathered skin depth. Thus, while the numerical value of $h$ remains consistent with \mbox{\citeA{Elliott2018}}, its physical interpretation is broadened to accommodate different surface processes. We also examine the effects of an impact on the thermal inertia of a material. The thermal inertia of a material is dictated by the thermal conductivity, $\kappa$, and the density of the material itself, $\rho$. \cref{eq: Thermal inertia} shows this relationship, with $\kappa$ in units of Wm$^{-1}$K$^{-1}$, $\rho$ in kg$\cdot$m$^{-3}$, and C, the material's specific heat capacity, is in units of J kg$^{-1}$K$^{-1}$.

\begin{equation}
    I=\sqrt{\kappa \rho C}
    \label{eq: Thermal inertia}
\end{equation}

As seen from \cref{eq: Thermal inertia}, any reduction in density or thermal conductivity leads to a decrease in the thermal inertia. Thus, we believe that Martian crater rays are formed by an increase in the porosity of the surface regolith immediately following an impact. This increased porosity decreases the density of the substance and lowers the thermal inertia.

\subsection{Martian Craters}

\begin{figure}
    \noindent\includegraphics[width=\textwidth]{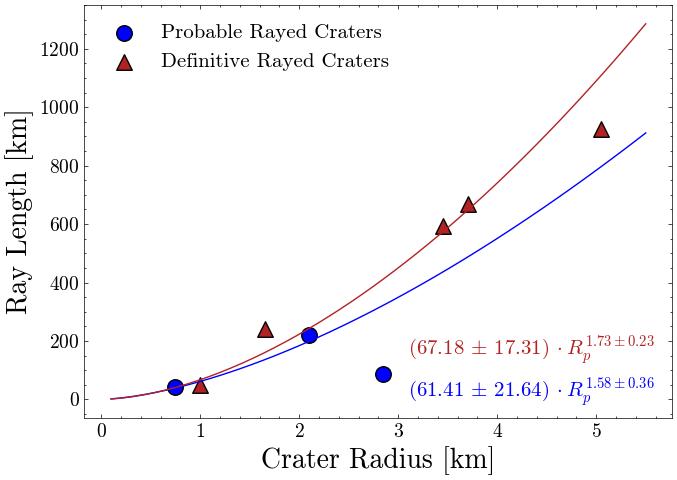}
    \caption{A comparison of the fit lines calculated by including the potential rayed craters described in \citeA{Tornabene2006}. The red line represents the fit line where only definitive rayed craters were included, and the blue line represents the fit when both definitive and probable rayed craters were included. While the value of $R_1$ in \mbox{\cref{Elliot_equation}} remains fairly similar, the value of $s$ changes by about 1.1x. Regardless, we chose to rely on the expertise of the researchers in Tornabene et al. (2006) and deferred to their assessment of these probable craters, ultimately removing them from our analysis, and using only four rays from Tornabene et al. (2006) and one ray from McEwen et al. (2005).}
    \label{ProbableCraters}
\end{figure}



A table showing the complete list of Martian craters identified by \citeA{Tornabene2006} and \citeA{McEwen2005} used in this work can be found in \cref{TornabeneCraters} of the appendix. Additional Martian rayed craters have since been cataloged in \citeA{Harris_2023}, but the length of the crater rays was not recorded, so in our work we only used data from \citeA{Tornabene2006} and \citeA{McEwen2005}. Of the seven craters presented in \citeA{Tornabene2006} four had definite crater rays and three had \textit{probable} rays. The inclusion of these probable crater rays was considered in our analysis and can be seen in \cref{ProbableCraters}. 

Although our Martian rayed crater dataset is limited in size, the craters included are among the freshest examples within their size class (as noted by \mbox{\citeA{Tornabene2006}}). Freshness is an important consideration because ray systems on Mars are thought to degrade relatively rapidly due to aeolian processes \mbox{\cite{Tornabene2006}}. Expanding the sample to include additional rays would likely introduce older craters whose rays have undergone substantial erosion or surface modification, systematically biasing the measured lengths toward shorter values. While broader mapping efforts would certainly be valuable for future studies, this potential degradation effect represents an additional factor that must be carefully considered when incorporating larger or more heterogeneous datasets.

The addition of the three probable crater rays modifies both the exponent $s$ and the leading constant $R_1$, in our calculated power-law fit. While the value of $R_1$ in \mbox{\cref{Elliot_equation}} remains fairly similar, the value of $s$ changes by about 1.1x. Regardless, we chose to rely on the expertise of the researchers in \mbox{\citeA{Tornabene2006}} and deferred to their assessment of these probable craters, ultimately removing them from our analysis, and using only four rays from \mbox{\citeA{Tornabene2006}} and one ray from \mbox{\citeA{McEwen2005}}.

\subsection{Fit Parameters}

 To use the model from \citeA{Elliott2018}, we need constrained the following values: $R_{ref}$, the radius of our reference crater; $d_{boulder}$, the minimum diameter of the boulders around the crater; $n_0$, the density of the boulders on the rim of the reference crater; $\rho$, the target density; $\delta$, the projectile density, $\beta$ and $\kappa$, both constants that describe the impacted surface; and $\alpha$, a scaling constant for the excavation depth. A full list of these values and their units is provided in \cref{Table:Parameters}. Some values, such as $\beta$ and $\kappa$, both constants for dry sand, remained unchanged. The remaining values were obtained from published literature and are discussed in detail below, and their ranges can be seen in \cref{Table:ValuesUsed}.

\subsubsection{Crater Size and Density}
Crater S1094b was chosen to be the reference crater used in the Martian analysis. This crater was created recently and recorded by NASA's InSight mission on December 24, 2021 as a seismic event of magnitude $>$4 \cite{Posiolova}. A hypervelocity impact formed crater S1094b, and it has a diameter of ~155 m and a depth of ~20-23m \cite{Posiolova}.  \citeA{Tusberti2024} mapped the boulder density around crater S1094b, providing an analog to the lunar study of Censorius crater by \citeA{Krishna2016}. Their work was also used to obtain values for parameters in \cref{Table:Parameters}, such as the size of the crater ($R_{ref}$=77.5m) and boulder diameters ($d_{boulder}$=1.5m). To refine the analysis, Tusberti categorized boulders into icy and rocky types, but only rocky boulders were considered in this study to align with \citeA{Elliott2018}. In addition, \citeA{Tusberti2024} measured the boulder density twice - first in February 2022 and then again in December 2022, finding different densities each time. February had a higher boulder density, with 2304 boulders/km$^2$ with $d_{boulder} \geq$ 1.50m. The density in December was significantly lower, with only 462 boulders/km$^2$, an 80\% loss in counts. Both extremes were used to test fits, however, they did not affect the final outcome, likely due to the weak dependence of $R_1$ on boulder density (see \cref{R1 equation}). The results shown in \cref{fig:SkinDepthCompare} use the February boulder density to more closely match the boulder density of 4327 boulders/km$^2$ found by \citeA{Krishna2016} around Censorinus crater on the Moon.

\subsubsection{Dependence on Crater Size}
Following the procedure given in \citeA{Elliott2018}, the values of $a$; ejecta fragment cumulative size frequency distribution (CSFD), $b$; and the excavation depth, $h$, were determined. By solving \cref{R1 equation} and \cref{s equation} for $R_1=67.18$ and $s=1.73$ respectively, \cref{fig:SkinDepthCompare} was created to show potential solutions to the fit by \citeA{Elliott2018}. The intersections of the dashed and solid lines represent points where both \cref{R1 equation} and \cref{s equation} are satisfied.

In the fit created by \citeA{Elliott2018}, the value $h$ represents the skin depth, or the depth of the mature lunar soil. A secondary impact needs to excavate this much soil at a minimum to bring bright material to the surface and contribute to creation of a ray. 
A value of $h=10$ cm means that a secondary impact would need to excavate at least 10 cm of regolith to form an optically visible ray (this is known as High Energy Deep Excavation, see \cref{Ejecta Types Elliott}). A secondary impactor that can only excavate 9 cm of material in that region would fall into the High Energy Shallow Excavation regime, and an optical ray would not be visible as the impactor would penetrate through the dark space-weathered surface material. However, the model by \citeA{Elliott2018} uses the $h$ parameter to only consider optical rays, and not thermal rays, which may form with excavation depths less than $h$. Since the surface regolith on Mars lacks a space-weathered surface, we consider $h$ the minimum depth of excavation needed to form a ray, and refer to it as the excavation depth (see section 2.1). We apply this definition to both thermal and optical rays in this work. 

Although larger values of $h$ were tested, the values shown in \cref{fig:SkinDepthCompare}b are the excavation depth values that intersected with the $c$ curves. Martian values of $h$ are only slightly smaller than those calculated by \mbox{\citeA{Elliott2018}} for lunar soil (mm to cm scale on Mars compared to $\sim$5-10 cm on the Moon). The surface of Mars experiences aeolian weathering, and if this was the true excavation depth, then the winds on Mars would overturn the young soil almost continuously. However, we believe that the excavation depth value may not represent the depth of material disruption, and we discuss this further in section 5. 

The values of both $c$ and $P$ are difficult to approximate, but have extremes that are easier to visualize. The dependence on primary crater size is represented by $c$. A value of $c=0$ would mean that the number of boulders would not depend on the size of the primary crater, and a value of $c=1$ would mean that the boulder population of the crater is linearly dependent on the size of the boulder. Values of $c$ ranging  from 0 to 1 were tested and shown in \cref{fig:SkinDepthCompare}. The value $P$ represents the probability (from 0 to 1) that a crater will excavate the surface to a depth $h$. In a scenario where $P=0$, there would be no overturn of the surface material, and no rays would form, since \cref{Elliot_equation} would equal zero. By contrast, a value of $P=1$ would mean that every piece of ejecta would excavate the surface to a depth of $h$ and be visible as bright unweathered secondary ejecta. \citeA{Elliott2018} assumes a value of $P=0.5$, but this work varied $P$ between $0.2$ and $0.8$ to ensure a more robust coverage. Results with $P$=0.5 are shown in \cref{fig:SkinDepthCompare}

\begin{figure}[h]
    \centering
    \includegraphics[width=0.6\linewidth]{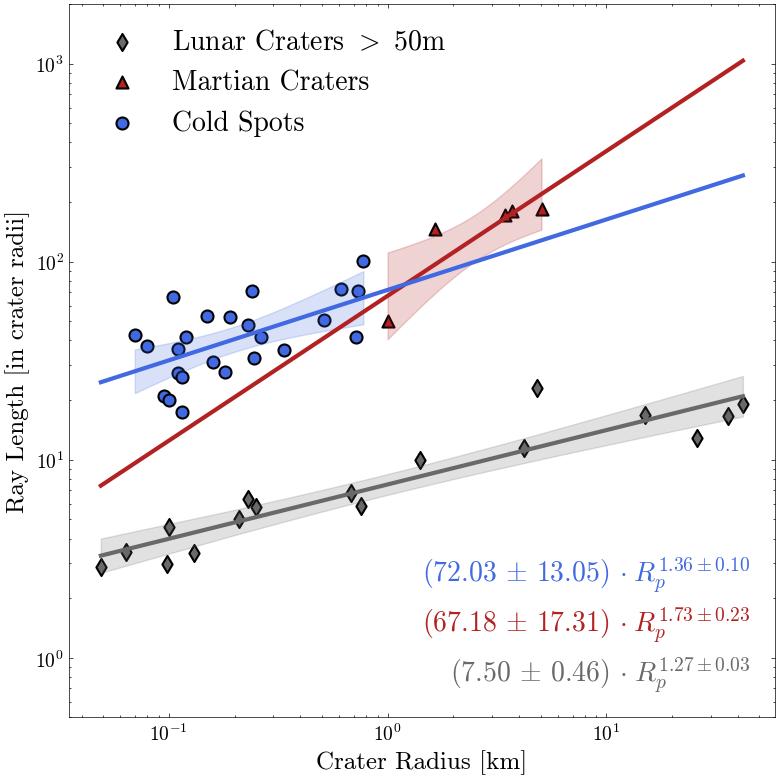}
    \caption{A comparison of crater ray length and the crater radius calculated for visible albedo lunar rays by \citeA{Elliott2018} (grey), Martian crater rays from \mbox{\citeA{McEwen2005}} and \mbox{\citeA{Tornabene2006}} (red), and lunar cold spots from \mbox{\citeA{Bandfield_2014}} (blue). Although many more rayed and cold spot craters have been detected, only a small fraction have had the length of their rays determined. Confidence intervals at 95\% confidence are shown for all three lines. The power-law index values for cold spots and albedo lunar rays are nearly identical, differing only slightly outside their respective error ranges. However, the leading constant, $R_1$, differs by more than an order of magnitude, with both Martian crater rays and cold spots exhibiting substantially larger values than albedo lunar rays. The leading constant for Martian rays falls within the error range of the value for lunar cold spots. Lunar craters from \mbox{\citeA{Elliott2018}} with diameters smaller than 50m were not used in this study. Some of these small craters have rays longer than predicted by the Elliott model and may represent bright ejecta from the primary crater deposited at the surface. Craters of similar sizes were not considered for Martian rays or lunar cold spots (see Section 3 for additional information).}
    \label{fig:compare_all}
\end{figure}

A comparison of the crater radii to ray lengths of Martian and lunar crater rays using \cref{Elliot_equation} can be seen in \cref{fig:compare_all}.

\section{Comparing Martian Rays and Lunar Cold Spots}

Despite having different proposed formation mechanisms, the physical properties of lunar cold spots and Martian crater rays are very similar. Both have radial ejecta patterns with low thermal inertia \cite{Hayne_2017,McEwen2005}, ejecta patterns visible with thermal imaging \cite{Bandfield_2014,Preblich2007}, and are expected to be a few centimeters thick \cite{Bandfield_2014,Hayne_2017,Tornabene2006}. One major difference they have is their estimated lifetimes. Lunar cold spots are only expected to last a few hundred thousand years \cite{Williams_2018}, whereas Martian crater rays have lifetimes on the order of tens of Mya \cite{McKeown2019,Harris_2023}. This is most likely explained by the impact gardening rate on the Moon being significantly larger than that on Mars \cite{Melosh1989}, which would allow surface features such as rays to exist for longer stretches of geological time on Mars. 

In this work, we use cold spot data from \mbox{\citeA{Bandfield_2014}}, Martian crater ray data from \mbox{\citeA{Tornabene2006}} and \mbox{\citeA{McEwen2005}}, and lunar crater data from \mbox{\citeA{Elliott2018}}. We acknowledge that our selection of cold spots and rayed Martian craters is relatively small, but out of the many more rayed and cold spot craters have been detected, only a small fraction have had the length of their rays determined. The focus of this work is on the comparison of rayed features on different bodies, and as a result, only values from the above publications were used in our analysis. 

Additionally, \mbox{\citeA{Elliott2018}} examined the behavior of rays around small craters ($<$ 50m in diameter) and found that they had rays longer than predicted by their model. They offered the explanation that these points are outliers produced primarily by deposition of low velocity ejecta from the primary crater itself rather than excavation of bright material by secondary impacts. These small craters are not expected to produce secondary craters at these distances. Section 4.3 of \mbox{\citeA{Elliott2018}} explains this in further detail. This explanation is further supported by \mbox{\citeA{Williams_2018}}, who found that the crater size-frequency distributions of smaller diameter craters exhibit a much shallower slope compared to medium or larger craters. They attributed this either to the difficulty of accurately measuring the lengths of small cold spots or to additional factors influencing cold spot formation at lower impact energies.  Since impact craters of comparable sizes are not present in our Martian or cold spots datasets, we excluded these small lunar craters from our analysis to prevent this end member behavior from disproportionately influencing only one dataset and biasing our cross comparison.

Analysis using the Elliott model of ray formation was performed on lunar cold spots as well, and compared to both lunar and Martian rays in \cref{fig:compare_all}. Despite having different values of $s$ and $R_1$, the majority of cold spot data points align with the Martian ray data. This may suggest that the true formation mechanism of both lunar cold spots and Martian crater rays are linked. 

\section{Excavation Depth and Depth of Disruption}
\begin{figure}
    \centering
    \includegraphics[width=0.85\linewidth]{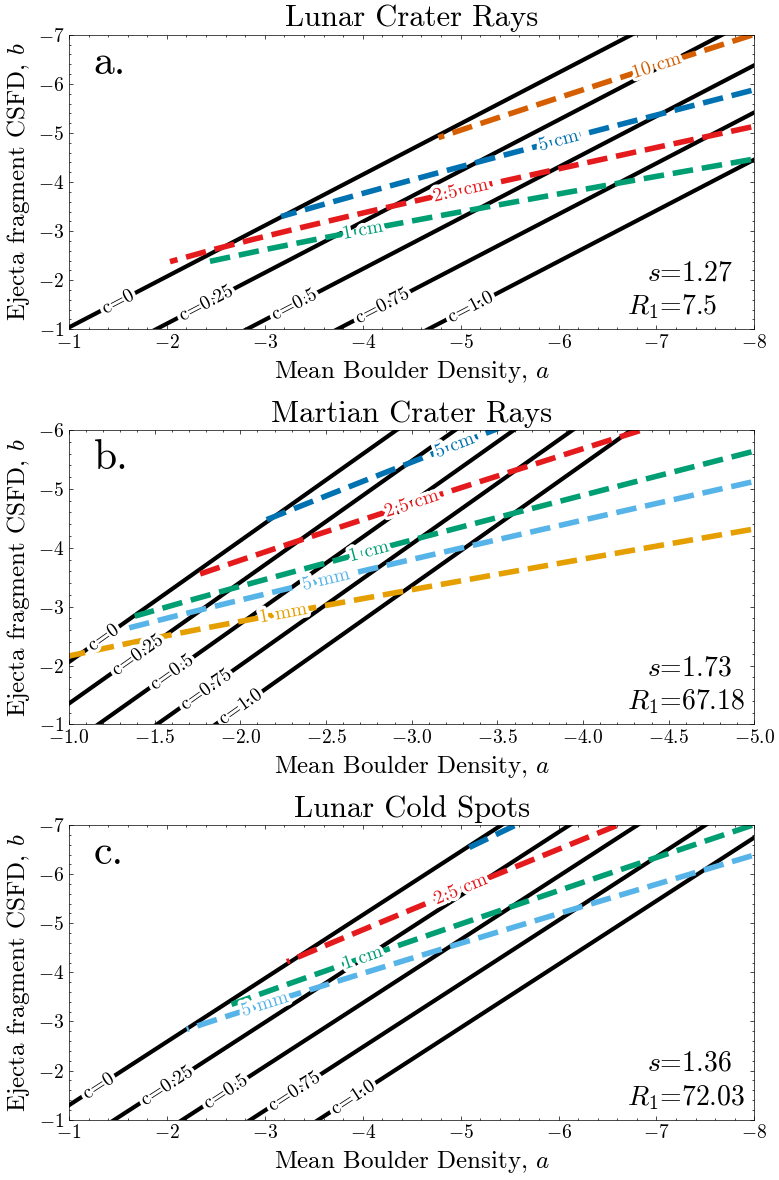}
    \caption{Plots of the minimum excavation depth needed to form a crater ray for different geologic features using the methodology in \citeA{Elliott2018}. The solid black lines are contours of $c$ that satisfy the value of $s$ written in the lower right. The dashed lines represent different values of $h$, the excavation depth, for the value of $R_1$ for each scenario. Intersection points of $h$ and the $c$ contours are values at which all components of the empirical fit derived by \citeA{Elliott2018} are satisfied. \textit{Note that $x$ and $y$ axes change in scale for figure 5b.} (a) A recreation of Figure 8 from \citeA{Elliott2018}, using the lunar parameters from \cref{Table:Parameters} derived by Elliott. (b) Excavation depth of the Martian crater rays from \citeA{Tornabene2006} and \citeA{McEwen2005}, using the Martian density parameters derived from literature (see \cref{Table:ValuesUsed}). (c) Excavation depth of the lunar cold spots from \citeA{Bandfield_2014} using the lunar parameters derived by \citeA{Elliott2018}.}
    \label{fig:SkinDepthCompare}
\end{figure}

To calculate the value of the excavation depth, \cref{Elliot_equation} was solved for the values of $a$, $b$, and $h$, by plugging in the calculated values of $s$ and $R_1$ shown in \cref{fig:compare_all} for each fit. The results of this parameterization can be seen in \cref{fig:SkinDepthCompare}

\citeA{Elliott2018} derived a value for the excavation depth of lunar rays on the order of 10 cm, which agrees with prior literature (see \citeA{Bandfield_2014,Hayne_2017}). However, applying this same model to lunar cold spots and Martian crater rays shows a much smaller excavation depth (on the order of 1 mm to 5 cm). While these excavation depths may seem low compared to albedo rays, it matches the expectation we would have for thermal inertia rays (see \mbox{\cref{Ejecta Types Elliott}}). As discussed in the following section, impact disruption can extend 10-100 times deeper than the depth of excavation.

Recent data from \citeA{Tusberti2024} shows that the boulder density around the Martian crater S1094b is $b=-4.68$ for the February distribution and $b=-3.47$ for the December distribution, both of which correspond to small excavation depth values ($h$). Similar work by \citeA{Button2017} on boulder distribution in the Tharsis Region and Elysium Planitia revealed a comparable density profile, with values of $b$ in the same regime ($b \approx -3$). 

To further investigate this, we tested boulder parameters from \citeA{Elliott2018} on Martian crater data and found that they only had a modest effect on the results (see \mbox{\cref{fig:ComparningLunarMartianParams}} in the Appendix). This suggests that the observed differences between lunar and Martian rays are not solely due to variations in boulder distributions, but rather point to a difference in the underlying physics behind the different types of ray creation. While Elliott’s framework accurately predicts the length and skin depth of visible albedo rays, thermal rays may exhibit distinct properties not accounted for in their model.

\section{Discussion}

The goal of this work was to compare the lengths of Martian rayed craters to those found on the Moon. Given the large difference in lunar and Martian $R_1$ values shown in \cref{fig:compare_all}, it seems that the function derived by \citeA{Elliott2018} may be missing some fundamental physics behind ray formation. The values of both the power law index, $s$, and the initial constant, $R_1$, increase in both the Martian and cold spot scenario, with the calculated Martian $R_1$ value being nearly 10x the size of the lunar $R_1$ given by Elliott. This is apparent in \cref{fig:compare_all} where lunar rays are depicted alongside Martian rays. A rayed crater with the same radius on both bodies would have rays nearly 10x longer on Mars compared to the Moon.

To determine if our chosen crater, S1094b, affected our results, we also tested the Elliott model on the Martian crater diameters and ray lengths using the lunar parameters that \mbox{\citeA{Elliott2018}} derived. This means we used a lunar crustal density, projectile density, and Censorinus crater on the Moon as the reference crater. Despite all these changes, we found that the excavation depth contours did not change by a significant amount. Further detail can be found in the Appendix.
 
Longer ray lengths on Mars initially seems to go against intuition, as Mars has a higher gravity than the Moon, and the presence of a thin atmosphere. However, as shown by \citeA{Elliott2018}, when assuming ballistic trajectories for crater rays, the gravity term itself cancels out and would not affect ray formation (see equations 4-6 in \citeA{Elliott2018}). Mars' thin atmosphere is also thought to have little effect on objects larger than 20 cm \cite{Artemieva_2004}, which is smaller than many secondary impactors.

In this work, lunar cold spots have been compared to both Martian crater rays and visible albedo lunar rays. Lunar cold spots and Martian rays have been shown to have similar features, including low thermal inertia \cite{Hayne_2017,Preblich2007} and, as this work found, both lunar cold spots and Martian thermal rays have lengths an order of magnitude longer than visible optical rays found around lunar craters (see \cref{fig:compare_all}). To further investigate this connection, we now examine how these features may share a common formation process.  

\begin{figure}
    \centering
    \includegraphics[width=0.85\linewidth]{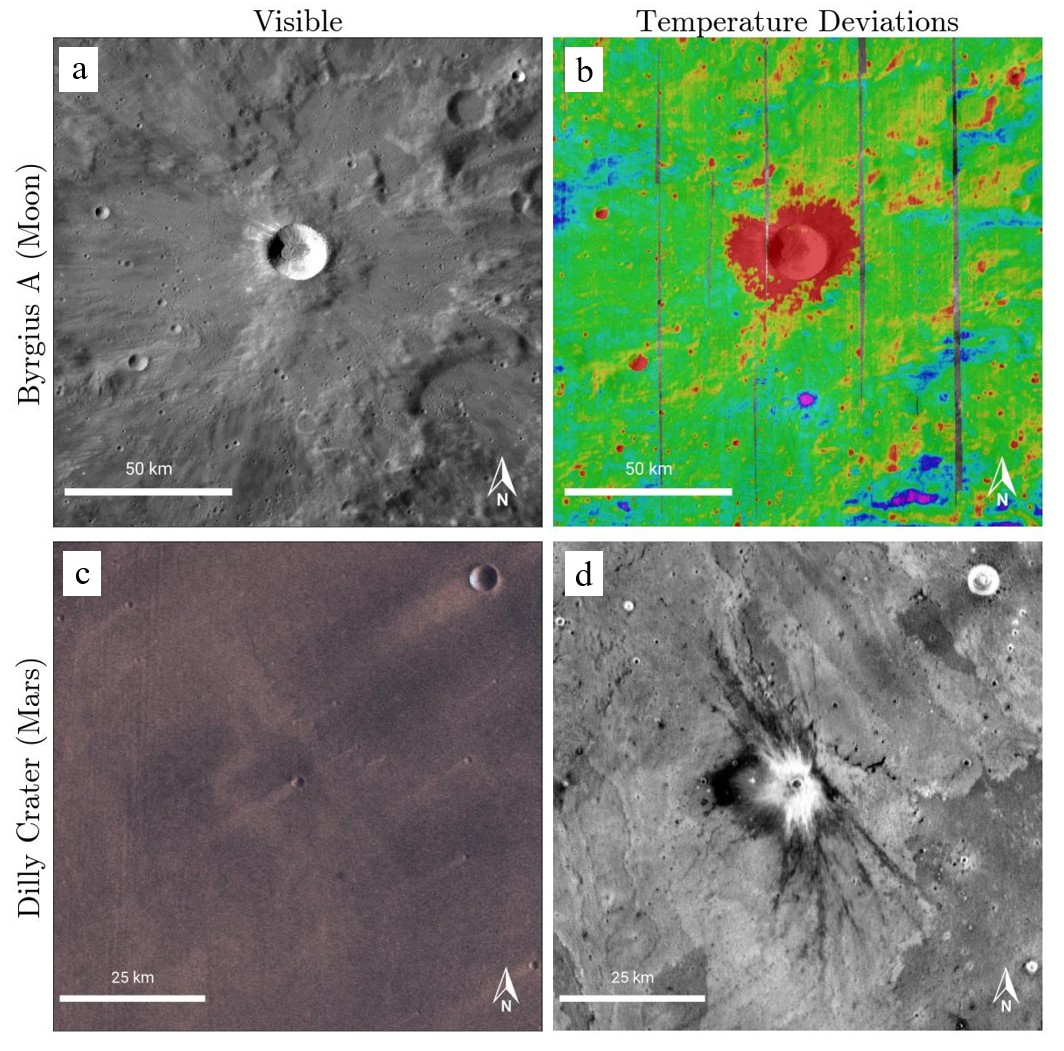}
    \caption{A comparison of thermophysical rays on the Moon and Mars. (a) Byrgius A, a lunar crater located at 24.55$^\circ$ S, 296.2$^\circ$ E with a diameter of 18.2km. (b) LRO Diviner Normalized Rock Free Regolith Temperature of Byrgius A showing warm, thermophysical rays \mbox{(red)} in and around the crater rim. (c) Dilly, a Martian crater located at 13.266$^\circ$ N, 157.22$^\circ$ E. (d) Nighttime temperature of Dilly crater using THEMIS, showing warm thermophysical rays (white) and colder thermal inertia rays (black).}
    \label{fig:GhentRays}
\end{figure}

Recent work by \mbox{\citeA{Ghent_2024}} discussed thermophysical rays on the Moon, which are warm anomalies that form around a rayed crater near the rim. Unlike cold spots and low thermal inertia spots, the surfaces of these thermophysical rays have \textit{increased} temperatures compared to the surrounding material. However, this may not be entirely unique to the Moon. \mbox{\citeA{Tornabene2006}} discusses similar warm anomalies with high thermal inertia around Martian rayed craters, and these thermophysical rays can even be seen around lunar cold spot craters (such as Bandfield crater, see \mbox{\cref{fig:Bandfield_ColdSpot}}) A depiction of this can be seen in \mbox{\cref{fig:GhentRays}}. Warm thermophysical rays such as those included in \mbox{\citeA{Harris_2023}} likely last longer than cold rays such as those found in \mbox{\citeA{Tornabene2006}}. To our knowledge, no prior publication has quantified the lengths of warm lunar thermophysical rays. As such, the ejecta behavior we propose should be viewed as a theoretical model intended to guide future observational work.

In our model, secondary cratering serves as the primary mechanism for the disruption of the surface regolith at long distances, contributing to the formation of both Martian crater rays and lunar cold spots. \cref{fig:SkinDepthCompare} shows that excavation depths for Martian crater rays and lunar cold spots are significantly smaller than those of albedo-based lunar rays. This fits with our understanding following the logic presented in \mbox{\cref{Ejecta Types Elliott}}. Prior work (\citeA{Wiggins_2019} and \citeA{Wiggins_2022}) has shown that hypervelocity impactors generate rarefaction waves, inducing porosity changes in materials up to 1-10 crater diameters away. Using the relationship between excavation depth and crater diameter from \citeA{Melosh1989}, these porosity changes would occur at 10-100 times the excavation depth. While \citeA{Wiggins_2019} and \citeA{Wiggins_2022} focused on primary hypervelocity impacts (10-20 km/s), the resulting disruption from the rarefaction wave extends far beyond the crater itself. This effect would likely persist in secondary impacts, which occur at much lower velocities ($<$1 km/s) \cite{Melosh1989}.

We can do a very conservative estimate using Martian crater rays. One of the calculated minimum excavation depth values was 1 cm. If we assume that this crater was formed via a secondary impact, and that it only affected surrounding material 2 crater radii away, that means that the depth of material disruption would be twenty times our original excavation depth, or a 20 cm area around the impact. This disrupted area would have a higher porosity than the surrounding material, and by \cref{eq: Thermal inertia}, a lower thermal inertia too. 

Martian crater rays are often found in regions of low to moderate albedo (0.10-0.26) and high background thermal inertia (110-330 J m$^{-2}$ K$^{-1}$ s$^{-1/2}$), defined as Thermophysical unit C by \citeA{Mellon_2000} and \citeA{Putzig_2005}. This region is believed to be duricrust, a formation consisting of hardened and solidified surface regolith \cite{Putzig_2005}. When an impactor strikes a duricrust-covered portion of the Martian surface, the impact would produce a rarefaction wave that breaks up the duricrust, causing the porosity to increase and the density to decrease \cite{Wiggins_2019,Wiggins_2022}. By \cref{eq: Thermal inertia}, this effect would cause a decrease in the thermal inertia, leading to the ray-like patterns of low thermal inertia seen on the Martian surface. While a majority of rayed craters are found within Unit C, a smaller subset is found within Unit B \mbox{\cite{Harris_2023}}, which, like Unit C, exhibits high background thermal inertia (160-355 J m$^{-2}$ K$^{-1}$ s$^{-1/2}$, \mbox{\citeA{Putzig_2005}}). This thermal contrast between the rays and their surroundings enables the low-inertia streaks to remain thermally distinct and thus detectable in orbital data.

On the Moon, surface regolith is often not as compact as Martian duricrust. However, micrometeorite bombardment has been known to cause small impact melts on the lunar surface \cite{Keller_1997,Wu_2017,Stojic_2023}. If these micrometeorite impacts were to be sufficient enough to create a thin layer of weak cementation over a period of geologic time, it would slowly increase the thermal inertia and conductivity. The rarefaction wave created by a large impactor would then be able to disrupt or modify this surface layer, reducing the thermal conductivity, which in turn would lower the surface's thermal inertia, creating the lunar cold spots we see today. We speculate that the formation of impact glass and agglutinates may increase thermal conductivity, and that subsequent disruption of this material may reduce it. It is also possible the typical state of the regolith is slightly compacted compared to newly disrupted regolith. \mbox{\citeA{FRIZZELL_2025}} hypothesize that impacts produce a surface dilating wave that can create the observed decompaction associated with cold spots. Their model indicates that the radial extent of this wave and how it corresponds to cold spot size depends on the layered structure of the Moon's shallow subsurface. It is unclear if this model could be extended to Martian rays, as many of these rays are found on young volcanic plains which have a more consolidated subsurface structure than the Moon. Thus, the work of \mbox{\citeA{FRIZZELL_2025}} may suggest that Martian rays have a distinct origin from lunar cold spots.

If what we propose is the true mechanism of thermal ray formation, then the High Energy Shallow Excavation regime depicted in \cref{Ejecta Types Elliott} represents the cutoff point for visible rays. The thermal rays visible on both the Moon and Mars would thus be just an extension of the visible rays we already can see.

\section{Conclusions}
To summarize the points made in the discussion section, this work has done the following:
\begin{enumerate}
    \item Shown that Martian crater rays are longer than lunar ones by an order of magnitude for craters with similar diameters (see \cref{fig:compare_all}).
    \item Demonstrated that Martian craters have ray lengths similar to those found around lunar cold spots (see \cref{fig:compare_all}).
    \item Proposed a relationship between the formation mechanisms of lunar cold spots and Martian crater rays
    \item Expanded upon the ray formation model by \citeA{Elliott2018} by including thermal rays in \cref{Ejecta Types Elliott}. 
    \item Proposed that Martian crater rays and lunar cold spots may be extensions of optically visible rays.

\end{enumerate}

Future work should further investigate the relationship between lunar cold spots and Martian crater rays. This work only used a small subset of Martian craters and lunar cold spots with publicly available ray lengths, but there are other cataloged rayed craters and cold spots that do not have recorded ray lengths (see \citeA{Harris_2023}). Including more Martian rayed craters and lunar cold spots may show a further correlation between these two geologic features. This also applies to warm, thermophysical rays. Future analysis of the lengths of these warm thermophysical rays could provide further insight into ejecta-surface interactions. Additionally, crater rays are not only exclusive to Mars and the Moon. Mercury also has a number of crater rays with similar physical properties to those on Mars. Incorporating Mercurian rays in a future study may provide further insights into thermal ray formation.

Updates to models such as the one by \citeA{Elliott2018} should be considered to more accurately predict the lengths of thermally visible crater rays. The point where rays are no longer optically visible may not mark the end of their formation, but rather the limit of their detectability by the naked eye.

\newpage
\section*{Open Research Section}
In this work, data from a number of prior papers was used for all fits. This includes lunar crater data from Table 2 of \citeA{Elliott2018}, Martian crater ray data from Table 1 of \citeA{Tornabene2006} (also as \cref{TornabeneCraters} in this paper), and lunar cold spot data from Table 1 of \citeA{Bandfield_2014}.  Figures were made with Matplotlib version 3.10.0 \cite{Caswell_2020,Hunter_2007}, available under the Matplotlib license at https://matplotlib.org/.

\acknowledgments
This work was supported by the NASA SSERVI Lunar Structure, Composition, Processes and Evolution (LunaSCOPE) Grant 80NSSC23M0161.
We would like to thank both of our reviewers for their insightful comments that helped us to improve the manuscript. 
Finally, special thanks go to Mia Rudin for her steadfast support and encouragement throughout this process. 

\section*{Conflict of Interest}

\noindent The authors declare no conflicts of interest.

\newpage
\bibliography{agusample}

\newpage
\section{Appendix} \hfill
\newline
\begin{table}[h]
    \centering
    \begin{tabular}{||c|c|c||}
    \hline
        Parameter & Units & Description \\
        \hline \hline
       $a$  & Unitless & Mean Spatial Density of Boulders \\
       $b$ & Unitless & Ejecta Fragment Cumulative Size Frequency Distribution \\
       $c$ & Unitless & Dependence on Crater Size\\
       $P$ & Unitless & Probability a surface is excavated to a depth $h$\\
       $\beta$ & Unitless &  Constant 0.17 for dry sand \cite{Schmidt1987}\\
       $k$ & Unitless & Empirical Scaling Constant 0.625 for Dry Sand \cite{Schmidt1987}\\
       $\alpha$ & Unitless & Maximum excavation depth of a streamline \cite{Gault1974} \\
       $h$ & km & Depth of Excavation \\
       R$_{ref}$ & km & Radius of the reference crater \\
       d$_{boulder}$ & km & Minimum Diameter of boulders around the Crater \\
       $n_0$ & Count per km$^2$ & Boulder Density at the Rim of the Reference Crater\\
       $\delta$ & kg/m$^3$ & Projectile Density \\
       $\rho$ & kg/m$^3$ & Target Density \\
       \hline
    \end{tabular}
    \caption[test]{A description of the parameters used in \citeA{Elliott2018}. Most values are unitless, but those with distance are all in km instead of m. \textit{Note: This is a necessary part of the fit derived by \cite{Elliott2018}. The function they derived in \cref{R1 equation} and \cref{s equation} only works with distance units in kilometers, not meters}. }
    \label{Table:Parameters}
\end{table}

\begin{table}[]
    \centering
    \begin{tabular}{||c|c|c||}
    \hline
        Parameter  & Values Tested & Source \\
        \hline \hline
       $a$   & -2 to -9 & Obtained from Fit \\
       $b$  & -2 to -9 & Obtained from Fit \\
       $c$  & 0 to 1 & \citeA{Elliott2018}\\
       $P$  & 0.2 to 0.8 & \citeA{Elliott2018}\\
       $h$ & 100 $\mu$m to 1 m & Obtained from Fit \\
       R$_{ref}$  &  77.5 m & \citeA{Tusberti2024} \\
       d$_{boulder}$ &  1.5 m & \citeA{Tusberti2024} \\
       $n_0$ & 462 to 2308 boulders/km$^2$ & \citeA{Tusberti2024}\\
       $\delta$ & 3100 kg/m$^3$  & \citeA{Baratoux2014}\\
       $\rho$ & 2582 kg/m$^3$   & \citeA{Goossens2017}\\
       \hline
    \end{tabular}
    \caption{An overview of the values tested while fitting the ray scaling law function given in  \citeA{Elliott2018} to the Martian craters from \citeA{McEwen2005} and \citeA{Tornabene2006}. The surface density of 2582 kg/m$^3$ is the crustal surface density of Mars, and the projectile density of 3100 kg/m$^3$ is the estimated density of surface boulders.}
    \label{Table:ValuesUsed}
\end{table}

\begin{table}[h]
    \centering
    \begin{tabular}{||c|ccccc||}
    \hline
        Crater Name & Radius [km] & Lat. & Long. & Median Ray Length [km] & Source\\
        \hline \hline
        Zunil & 5.05 & 7.70 & 166.00 & 927 &\citeA{McEwen2005}\\
        Tomini & 3.70 & 16.27 & 125.90 & 668 &\citeA{Tornabene2006}\\
        Gratteri & 3.45 & -17.70 & 199.9 & 595&\citeA{Tornabene2006}\\
        Zumba & 1.65 & -28.65 & 226.90 & 240&\citeA{Tornabene2006}\\
        Dilly & 1.00 & 13.27 & 157.23 & 50&\citeA{Tornabene2006}\\
        \textit{Tomini B*} & 2.10 & 14.90 & 123.25 & 220&\citeA{Tornabene2006}\\
        \textit{Crater A*} & 2.85 & 18.10 & 155.50 & 86&\citeA{Tornabene2006}\\
        \textit{Crater B*} & 0.75 & 15.50 & 159.20 & 42&\citeA{Tornabene2006}\\
        \hline
    \end{tabular}
    \caption{This table shows the Martian craters used in this study. The data on the craters themselves is taken from \citeA{Tornabene2006}, except for crater Zunil, which is taken from \citeA{McEwen2005}. A total of eight rayed craters can be found on Mars, but only the first five were used in this study. The last three craters, Tomini B, Crater A, and Crater B, were determined as only \textit{probable} rayed craters by \citeA{Tornabene2006}, so they were not included in this analysis.}
    \label{TornabeneCraters}
\end{table}

\begin{figure}
    \centering
    \includegraphics[width=0.98\linewidth]{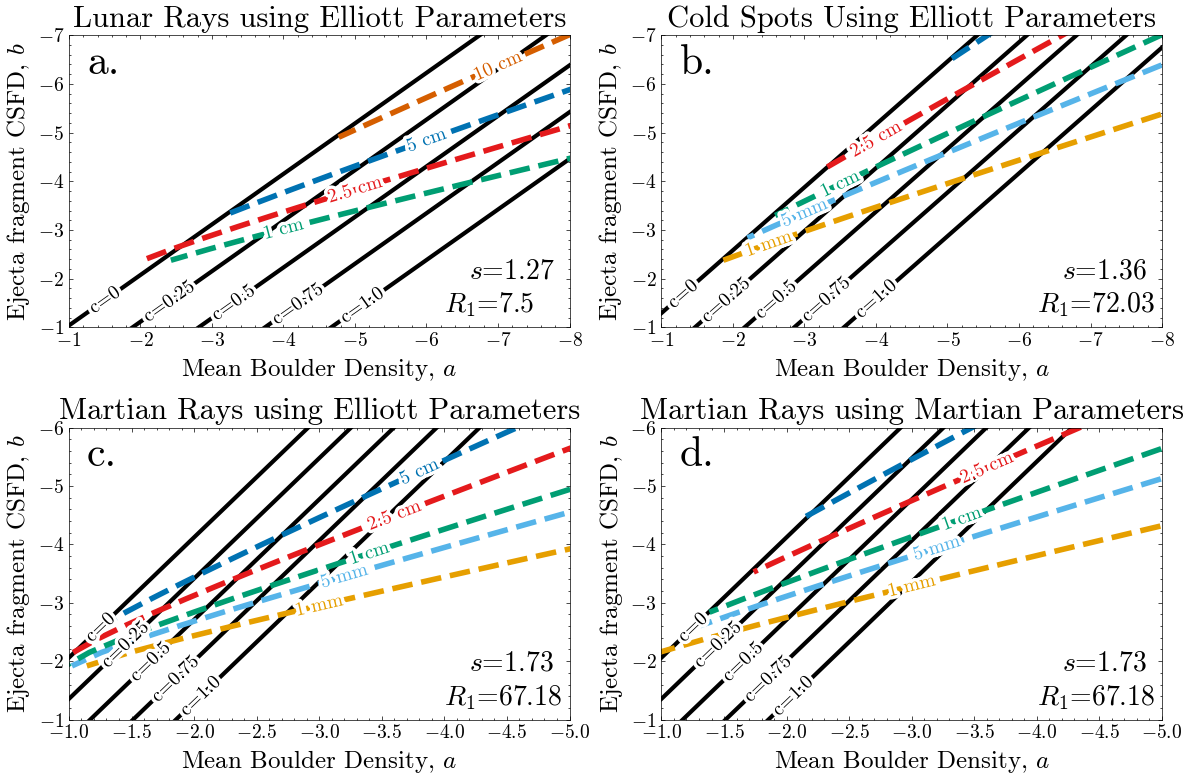}
    \caption{A comparison of lunar rays, cold spots, and Martian rays, both with Lunar parameters from \citeA{Elliott2018} and Martian parameters shown in \cref{Table:ValuesUsed}. It should be noted that \cref{fig:ComparningLunarMartianParams}a, b, and d correspond to the plots in \cref{fig:SkinDepthCompare}a, c, and b respectively. (a) A recreation of Figure 8 from \citeA{Elliott2018} using the lunar parameters from Table 1 derived by Elliott. These values are specific to the Moon, and include crustal density, projectile density, reference crater size, and boulders around the reference crater. (b) Excavation depth of the lunar cold spots from \citeA{Bandfield_2014} using the lunar parameters derived from \citeA{Elliott2018} (c)  Excavation depth of the Martian crater rays from \citeA{Tornabene2006} and \citeA{McEwen2005}, using the lunar parameters derived by Elliott. (d) Excavation depth of the Martian crater rays from \citeA{Tornabene2006} and \citeA{McEwen2005}, using the Martian density parameters derived from literature (see \cref{Table:ValuesUsed}). Statistical t-tests were performed comparing the skin depth contours of \cref{fig:ComparningLunarMartianParams}c and \cref{fig:ComparningLunarMartianParams}d to determine if the effect seen was a byproduct of the specific crater chosen in this work. Skin depths of 5cm and greater were statistically significant, but the smaller skin depths (of 2.5cm and below) were determined to be not statistically different.}
    \label{fig:ComparningLunarMartianParams}
\end{figure}

\begin{figure}
    \centering
    \includegraphics[width=0.98\linewidth]{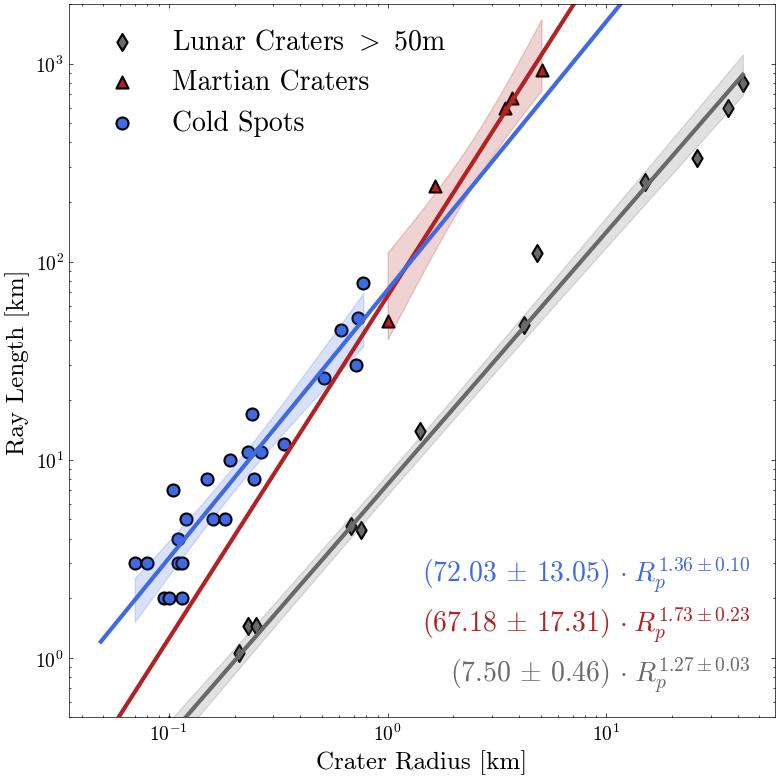}
    \caption{\textit{Note: This is a recreation of Figure 4, but with the y-axis scaled to crater ray length in km instead of crater radii.} A comparison of crater ray length and the crater radius calculated for visible albedo lunar rays by \citeA{Elliott2018} (grey), Martian crater rays from \mbox{\citeA{McEwen2005}} and \mbox{\citeA{Tornabene2006}} (red), and lunar cold spots from \mbox{\citeA{Bandfield_2014}} (blue). Lunar craters from \mbox{\citeA{Elliott2018}} with diameters smaller than 50m were not used in this study. Some of these small craters have rays longer than predicted by the Elliott model and may represent bright ejecta from the primary crater deposited at the surface. Craters of similar sizes were not considered for Martian rays or lunar cold spots (see Section 3 for additional information).}
    \label{fig:placeholder}
\end{figure}

\end{document}